\documentclass[12pt]{iopart}
\usepackage{graphicx}
\expandafter\let\csname equation*\endcsname\relax
\expandafter\let\csname endequation*\endcsname\relax
\usepackage{amsmath}
\usepackage{mathtools}
\DeclarePairedDelimiter\abs{\lvert}{\rvert}%
\DeclarePairedDelimiter\norm{\lVert}{\rVert}%
\usepackage{placeins}
\usepackage{textcomp}
\usepackage{url}

\allowdisplaybreaks

\let\Oldsection\section
\renewcommand{\section}{\FloatBarrier\Oldsection}

\let\Oldsubsection\subsection
\renewcommand{\subsection}{\FloatBarrier\Oldsubsection}

\let\Oldsubsubsection\subsubsection
\renewcommand{\subsubsection}{\FloatBarrier\Oldsubsubsection}

\makeatletter
\let\oldabs\abs
\def\abs{\@ifstar{\oldabs}{\oldabs*}}
\let\oldnorm\norm
\def\norm{\@ifstar{\oldnorm}{\oldnorm*}}
\makeatother

\makeatletter
\newcommand*{\centerfloat}{%
  \parindent \z@
  \leftskip \z@ \@plus 1fil \@minus \textwidth
  \rightskip\leftskip
  \parfillskip \z@skip}
\makeatother

\newcount\colveccount
\newcommand*\colvec[1]{
        \global\colveccount#1
        \begin{pmatrix}
        \colvecnext
}
\def\colvecnext#1{
        #1
        \global\advance\colveccount-1
        \ifnum\colveccount>0
                \\
                \expandafter\colvecnext
        \else
                \end{pmatrix}
        \fi
}

\begin{document}

\title[Indistinguishable quantum walks on graphs]{Indistinguishable quantum walks on graphs relative to a bipartite quantum walker}

\author{Phillip R. Dukes}

\address{University of Texas Rio Grande Valley, Brownsville, TX 78520, USA}
\address{26 October 2016}
\ead{phillip.dukes@utrgv.edu}

\begin{abstract}
A distinguishability operator is defined for the continuous-time quantum walk (CTQW) of a bipartite quantum walker on two simply connected graphs, $W_{G_i,G_j} = U_{G_i}\left(t\right) \otimes U_{G_j}\left(t'\right) -  U_{G_j}\left(t'\right) \otimes U_{G_i}\left(t\right)$, where $U_{G_i}\left(t\right)$ is the unitary CTQW operator for a labeled graph $G_i$ over a time interval $t$. The null space of $W_{G_i,G_j}$ defines the vector space of initial bipartite states whose time development is either constant or only dependent on $t + t'$ and is invariant to which quantum walker subsystem goes with each graph. The set of null spaces corresponding with a set of $W_{G_i,G_j}$ have interesting relations as subspaces, intersections between subspaces, and subspaces of intersections. These relations are depicted as Euler diagrams for labeled graphs of three and four vertices.
\end{abstract}

\section{Introduction}
Let $G=\left(V,E\right)$ be a simply connected and undirected labeled graph, where $V$ is a set of $n$ vertices and $E$ is a set of connecting edges. The adjacency matrix $A$ and degree matrix $D$ which describe the graph are defined as follows:
\begin{subequations}
\begin{align}
A_{j,k} &= \begin{cases} 1 &\mbox{if there is an edge connecting vertices \textit{j} and \textit{k}} \\
0 & \mbox{otherwise} \end{cases} \label{eq:Amat}\\
D_{j,k} &= \begin{cases} deg\left( v_{j}\right)  &\mbox{if }j=k \\
0 & \mbox{otherwise} \end{cases} \label{eq:Dmat}
\end{align}
\end{subequations}
where $deg\left( v_{j}\right)$ is the degree of vertex $v_{j}$.

 Generally, the unitary CTQW operator $U_G(t)$ is defined in terms of the Laplacian matrix $L$ of the graph $G$. \cite{Farhi:1, Gerhardt:1}
\begin{subequations}
\begin{align}
L &= D - A \label{eq:Laplacian}\\
U_G\left( t \right) &=e^{-i t L}. \label{eq:Uop}
\end{align}
\end{subequations}
The quantum walker is described by a time-dependent quantum state $\lvert \phi\left( t\right)  \rangle$. The initial state vector $\lvert \phi\left( 0\right)  \rangle$ is an ordered set of components, each component corresponding to the initial amplitude at each vertex of the graph (we will label each vertex starting with $0$)

\begin{equation}
\lvert \phi\left( 0\right)  \rangle  =\left(\phi_{0}\left( 0\right) ,\phi_{1}\left( 0\right),\phi_{2}\left( 0\right),...,\phi_{n-1}\left( 0\right)\right) 
\label{eq:initialState}
\end{equation}
such that
\begin{equation}
\langle\phi\left( 0\right)  \lvert \phi\left( 0\right)  \rangle  =  \displaystyle\sum_{j=0}^{n-1} \abs{\phi_{j}\left( 0\right)}^2 = 1 .
\label{eq:initialNorm}
\end{equation}
The time development of the quantum walk then becomes
\begin{equation}
U_G\left(t\right)\lvert \phi\left(0\right)  \rangle  = \lvert \phi\left(t\right)  \rangle,
\label{eq:QW}
\end{equation}
and the time-dependent amplitude at vertex $j$ is 
\begin{equation}
 \phi_{j}\left( t\right)  = \langle j \lvert \phi\left( t\right)  \rangle
\label{eq:amp-at-j}
\end{equation}
such that after a time evolution of $t=t_f$ the probability for observing the walker at vertex j is $\langle \phi_j\left( t_f\right) \lvert \phi_j\left( t_f\right) \rangle$.

\section{The bipartite quantum walker on two graphs}
For two quantum systems labeled $A$ and $B$ let $\mathcal{H}_A$ and $\mathcal{H}_B$ represent the Hilbert spaces for the state vectors $\lvert\psi\rangle_A$ and $\lvert\psi\rangle_B$ respectively. The Hilbert space for the composite system $A \cup B$ is then $\mathcal{H}_{AB} = \mathcal{H}_A \otimes \mathcal{H}_B$ and the composite state vector becomes
\begin{subequations}
\begin{align}
\lvert\psi\rangle_{AB} &= \displaystyle\sum_{i,j} \alpha_{i,j} \lvert i \rangle_A \otimes \lvert j \rangle_B, \label{eq:bipartite}\\
\displaystyle\sum_{i,j} \lvert \alpha_{i,j} \rvert^2 &= 1, \label{eq:norm}
\end{align}
\end{subequations}
where $\lvert i \rangle_A$ and $\vert j \rangle_B$ are standard bases in $\mathcal{H}_A$ and $\mathcal{H}_B$ respectively.

A bipartite state vector which can be represented as $\lvert \psi \rangle_{AB} = \lvert \psi \rangle_A \otimes \lvert \psi \rangle_B$ is called a separable state, otherwise it is called an entangled state. Unitary operations which are separable on the subsystems $A$ and $B$ are of the form $U_{AB} = U_A \otimes U_B$.

Given two graphs $G_1=\left( V_1,E_1\right)$ and $G_2=\left( V_2,E_2\right)$ the composite CTQW operator will be
\begin{equation}
U_{G_1,G_2}\left( t,t'\right) = U_{G_1}\left( t\right) \otimes U_{G_2}\left( t'\right),
\label{eq:compopp}
\end{equation}
where the quantum walk on graph $G_1$ ($G_2$) is for a time $t$ ($t'$). Accordingly, the time evolution of an initial bipartite walker will be
\begin{equation}
U_{G_1}\left( t\right) \otimes U_{G_2}\left( t'\right) \lvert\psi_0\rangle_{AB} = \lvert\psi\left( t,t'\right) \rangle_{AB},
\label{eq:compevol}
\end{equation}
in which $U_{G_1}\left( t_1\right)$ acts on subsystem $A$ and $U_{G_2}\left( t_2\right)$ acts on subsystem $B$ and $\lvert \alpha_{i,j} \rvert^2$ from equations \ref{eq:bipartite} and \ref{eq:norm} is the joint probability for finding subsystems $A$ and $B$ at vertices  $i$ and $j$ of graphs $G_1$ and $G_2$ respectively. The entropy of entanglement of $\lvert\psi_0\rangle_{AB}$ is invariant to the locally separable operator $U_{G_1,G_2}\left( t,t'\right)$ \cite{Vedral:1}.

\section{Indistinguishability relative to $\lvert\psi_0\rangle_{AB}$}
For the bipartite walker on two graphs $G_i$ and $G_j$ we define a non-unitary distinguishability operator 
\begin{equation}
W_{G_i,G_j} = U_{G_i}\left(t\right) \otimes U_{G_j}\left(t'\right) -  U_{G_j}\left(t'\right) \otimes U_{G_i}\left(t\right).
\label{eq:Wij}
\end{equation}
We are interested in the null space of $W_{G_i,G_j}$, i.e., the space of walker initial states which satisfy
\begin{equation}
W_{G_i,G_j}\lvert\psi_0\rangle_{AB} = 0
\label{eq:nullsts}
\end{equation}
with $t$ and $t'$ greater than $0$. Any initial state $\lvert\psi_0\rangle_{AB}$ in the null space of $W_{G_i,G_j}$ will have the property
\begin{equation}
U_{G_i}\left(t\right) \otimes U_{G_j}\left(t'\right) \lvert\psi_0\rangle_{AB} =  U_{G_j}\left(t'\right) \otimes U_{G_i}\left(t\right) \lvert\psi_0\rangle_{AB} = \lvert\psi\left( t+t'\right) \rangle_{AB}.
\label{eq:Wpsi}
\end{equation}
We point out that for any two graphs $G_i$ and $G_j$ a trivial solution to Equation \ref{eq:nullsts} (in addition to the zero vector) is the uniform amplitude $\lvert\psi_0\rangle_{AB}=\dfrac{1}{\sqrt{d}}\left( 1, 1,...,1\right)$ where $d$ is the cardinality of $\mathcal{H}_{AB}$,  $d = \lvert V_{G_i}\rvert \times \lvert V_{G_j} \rvert$. 

We will consider the condition when $\lvert V_{G_i}\rvert = \lvert V_{G_j} \rvert$: the two simply connected graphs have equal order $n$. For the special case $i=j$ the two graphs are isomorphic and are labeled equivalently although the duration of the quantum walk may be different for each graph. In this case 
\begin{equation}
\begin{aligned}
\mathrm{null}\left( W_{G_i,G_i}\right) &= \mathrm{null}\left( U_{G_i}\left(t\right) \otimes I_n -  I_n \otimes U_{G_i}\left(t\right)\right) = \mathrm{null}\left(L_{G_i} \otimes I_n -  I_n \otimes L_{G_i}\right) \\
 &= \mathrm{null}\left(I_n \otimes U_{G_i}\left(t'\right) -  U_{G_i}\left(t'\right) \otimes I_n\right) = \mathrm{null}\left(I_n \otimes L_{G_i} -  L_{G_i} \otimes I_n\right),
\end{aligned}
\label{eq:nullspac}
\end{equation}
where $I_n$ is the $n\times n$ identity matrix and $L_{G_i}$ is the Laplacian matrix of $G_i$ .

Solutions to \ref{eq:nullspac} are more easily obtained through analysis of an equivalent expression: \cite{Simoncini:1, Higham:1}.
\begin{equation}
\begin{aligned}
&\left(I_n \otimes L_{G_i} -  L_{G_i}\otimes I_n\right)\lvert\psi_0\rangle_{AB} = 0\\
&\mathrm{is\: equlivalent\: to}\\
&L_{G_i} X - X L_{G_i}=0,
\end{aligned}
\label{eq:nullmateq}
\end{equation}
where $X$ is a variable $n \times n$ matrix such that $\lvert\psi_0\rangle_{AB} = \mathrm{vec}(X)$, and every $L_{G_i}$ is symmetric.

When $i\neq j$, $G_i$ and $G_j$ may be non-isomorphic graphs or isomorphic graphs with different labeling. In either case the null space of $W_{G_i,G_j}$ will be
\begin{equation}
\mathrm{null}\left( W_{G_i,G_j}\right)\subseteq \mathrm{null}\left( W_{G_i,G_i}\right)\cap\mathrm{null}\left( W_{G_j,G_j}\right),
\label{eq:nullij}
\end{equation}
from which $\mathrm{null}\left( W_{G_i,G_j}\right)$ can be determined by inspection.

The null space of $W_{Kn,Kn}$, where $Kn$ is the complete graph of order $n$, encompasses all other $\mathrm{null}\left(W_{G_i,G_j}\right)$,
\begin{equation}
\mathrm{null}\left( W_{G_i,G_j}\right)\subseteq \mathrm{null}\left( W_{Kn,Kn}\right),
\label{eq:nullKK}
\end{equation} 
and 
\begin{equation}
\mathrm{null}\left( W_{G_i,G_i}\right) = \mathrm{null}\left( W_{G_i,Kn}\right).
\label{eq:nullKi}
\end{equation}

We find that it is convenient to work with and describe the null space of $W_{G_i,G_j}$ in a non-orthonormal basis in which the simple sum of the basis vectors is the uniform vector $\left( 1,1,...,1\right)$. In the next section we illustrate how this works.

\subsection{$\mathrm{null}\left( W_{G_i,G_j}\right)$ for graphs of order 3}
\begin{figure}[!htbp]
\centering
\includegraphics[scale=0.6]{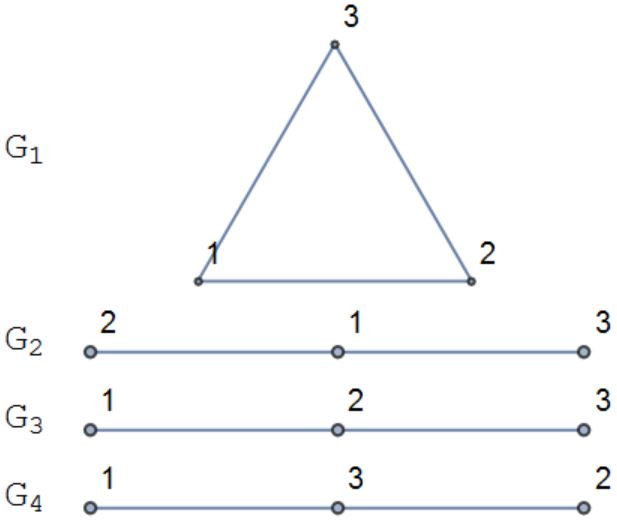}
\caption{All labeled simply connected graphs of order 3.}
\label{fig:g3s}
\end{figure}

There are 4 different labeled graphs with order 3, see Fig. \ref{fig:g3s}. 
By equations \ref{eq:nullmateq} through \ref{eq:nullKi} we obtain:

\begin{subequations}
\begin{align}
\underline{ W_{G_1,G_1}} &\rightarrow \begin{cases} \mathrm{vec}\left( X\right) =& \{ x_1, x_2, x_3, x_4, x_5, x_1 + x_2 + x_3 - x_4 - x_5,\\
&x_2 + x_3 - x_4,x_1 + x_3 - x_5,-x_3 + x_4 + x_5\} \\
\mbox{null basis:} & \{1, 0, 0, 0, 0, 1, 0, 1, 0\}\\
 & \{0, 1, 0, 0, 0, 1, 1, 0, 0\}\\
 & \{0, 0, 1,0, 0, 1, 1, 1, -1\}\\
 & \{0, 0, 0, 1, 0, -1, -1, 0, 1\}\\
 & \{0, 0, 0, 0, 1, -1, 0, -1, 1\} \end{cases} \label{eq:W11}\\
\underline{ W_{G_2,G_2},W_{G_2,G_1}} &\rightarrow \begin{cases} \mathrm{vec}\left( X\right) =& \{x_1, x_2, x_2,x_2, x_3, x_1 + x_2 - x_3, x_2,\\
 & x_1 + x_2 - x_3, x_3\} \\
\mbox{null basis:} & \{1, 0, 0, 0, 0, 1, 0, 1, 0\}\\
 & \{0, 1, 1, 1, 0, 1, 1, 1, 0\}\\
 & \{0, 0, 0, 0, 1, -1, 0, -1, 1\} \end{cases} \label{eq:W22}\\
\underline{ W_{G_3,G_3},W_{G_3,G_1}} &\rightarrow \begin{cases} \mathrm{vec}\left( X\right) =& \{x_1, x_2, x_3, x_2, x_1 - x_2 + x_3, x_2, x_3, x_2, x_1\} \\
\mbox{null basis:} & \{1, 0, 0, 0, 1, 0, 0, 0, 1\}\\
 & \{0, 1, 0, 1, -1, 1, 0, 1, 0\}\\
 & \{0, 0, 1, 0, 1, 0, 1, 0, 0\} \end{cases} \label{eq:W33}\\
\underline{ W_{G_4,G_4},W_{G_4,G_1}} &\rightarrow \begin{cases} \mathrm{vec}\left( X\right) =& \{x_1, x_2, x_3, x_2, x_1, x_3, x_3, x_3, x_1 + x_2 - x_3\} \\
\mbox{null basis:} & \{1, 0, 0, 0, 1, 0, 0, 0, 1\}\\
 & \{0, 1, 0, 1, 0, 0, 0, 0, 1\}\\
 & \{0, 0, 1, 0, 0, 1, 1, 1, -1\} \end{cases} \label{eq:W44}\\
 \begin{array}{l}
         \underline{ W_{G_2,G_3},W_{G_2,G_4},}\\
        \underline{ W_{G_3,G_4}}\end{array} &\rightarrow \begin{cases} \mathrm{vec}\left( X\right) =& \{x_1, x_1, x_1, x_1, x_1, x_1, x_1, x_1, x_1\} \\
\mbox{null basis:} & \{1, 1, 1, 1, 1, 1, 1, 1, 1\} \end{cases} \label{eq:W12}
\end{align}
\end{subequations}

We find the null spaces of the operators $W_{G_2,G_3},W_{G_2,G_4},$ and $W_{G_3,G_4}$ in equation \ref{eq:W12} according to equation \ref{eq:nullij}
\begin{equation}
\begin{aligned}
&\mathrm{null}\left( W_{G_2,G_2}\right)\cap\mathrm{null}\left( W_{G_3,G_3}\right)=\mathrm{null}\left( W_{G_2,G_2}\right)\cap\mathrm{null}\left( W_{G_4,G_4}\right)\\
&=\mathrm{null}\left( W_{G_3,G_3}\right)\cap\mathrm{null}\left( W_{G_4,G_4}\right)\\
&=\{\{0, 1, 1, 1, 0, 1, 1, 1, 0\}, \{1, 0, 0, 0, 1, 0, 0, 0, 1\}\},
\end{aligned}
\label{eq:nullijes}
\end{equation}
from which we determine 
\begin{equation}
\begin{aligned}
&\mathrm{null}\left( W_{G_2,G_3}\right)=\mathrm{null}\left( W_{G_2,G_4}\right)=\mathrm{null}\left( W_{G_3,G_4}\right)\\
&=\{1, 1, 1, 1, 1, 1, 1, 1, 1\}
\end{aligned}
\label{eq:nullijs}
\end{equation}
by inspection.

\begin{figure}[!htbp]
\centering
\includegraphics[scale=0.6]{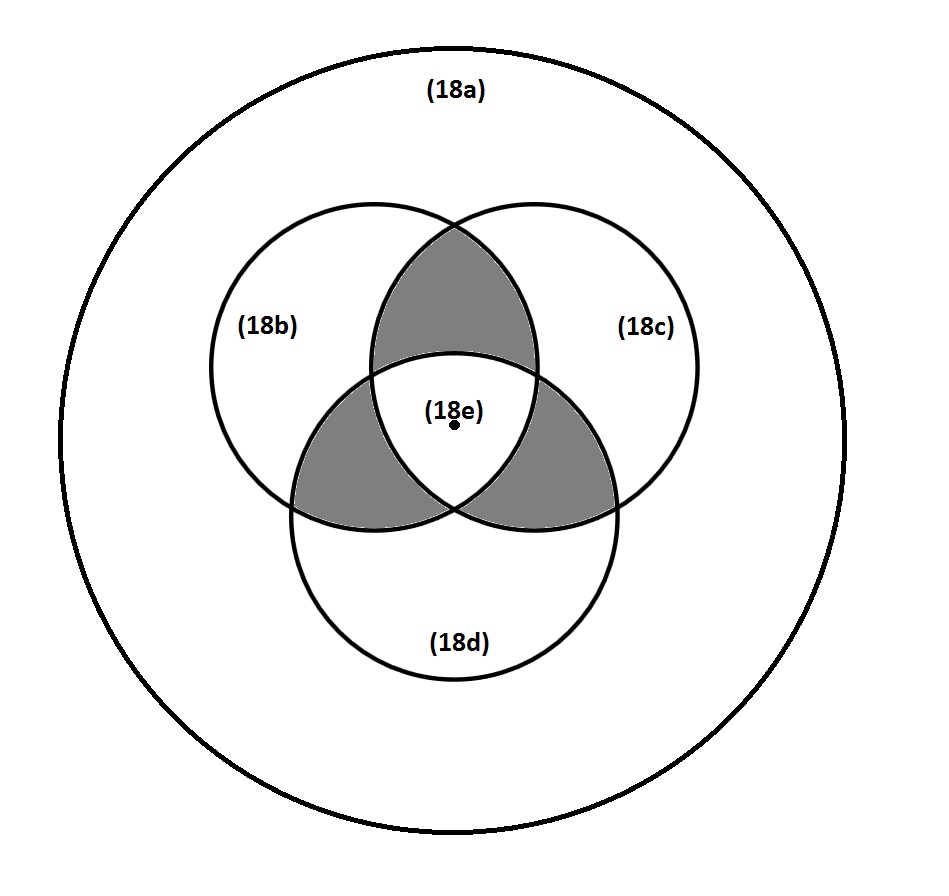}
\caption{Euler diagram of all $\mathrm{null}\left( W_{G_i,G_j}\right)$ for simply connected graphs of order 3. Each subspace/zone is annotated by the applicable equation number above. The shaded zones are empty.}
\label{fig:euler3}
\end{figure}

These results and their interrelationships can be represented as an Euler diagram in which closed curves and intersecting zones represent a null space $\mathrm{null}\left( W_{G_i,G_j}\right)$. As a mater of convenience, our Euler diagrams are non-area-proportional and wellformed up to labeling, sometimes having non-unique labeled curves \cite{Stapleton:1} (see Fig. \ref{fig:euler3}).  

\subsection{$\mathrm{null}\left( W_{G_i,G_j}\right)$ for graphs of order 4}
\begin{figure}[!htbp]
\centering
\includegraphics[scale=0.6]{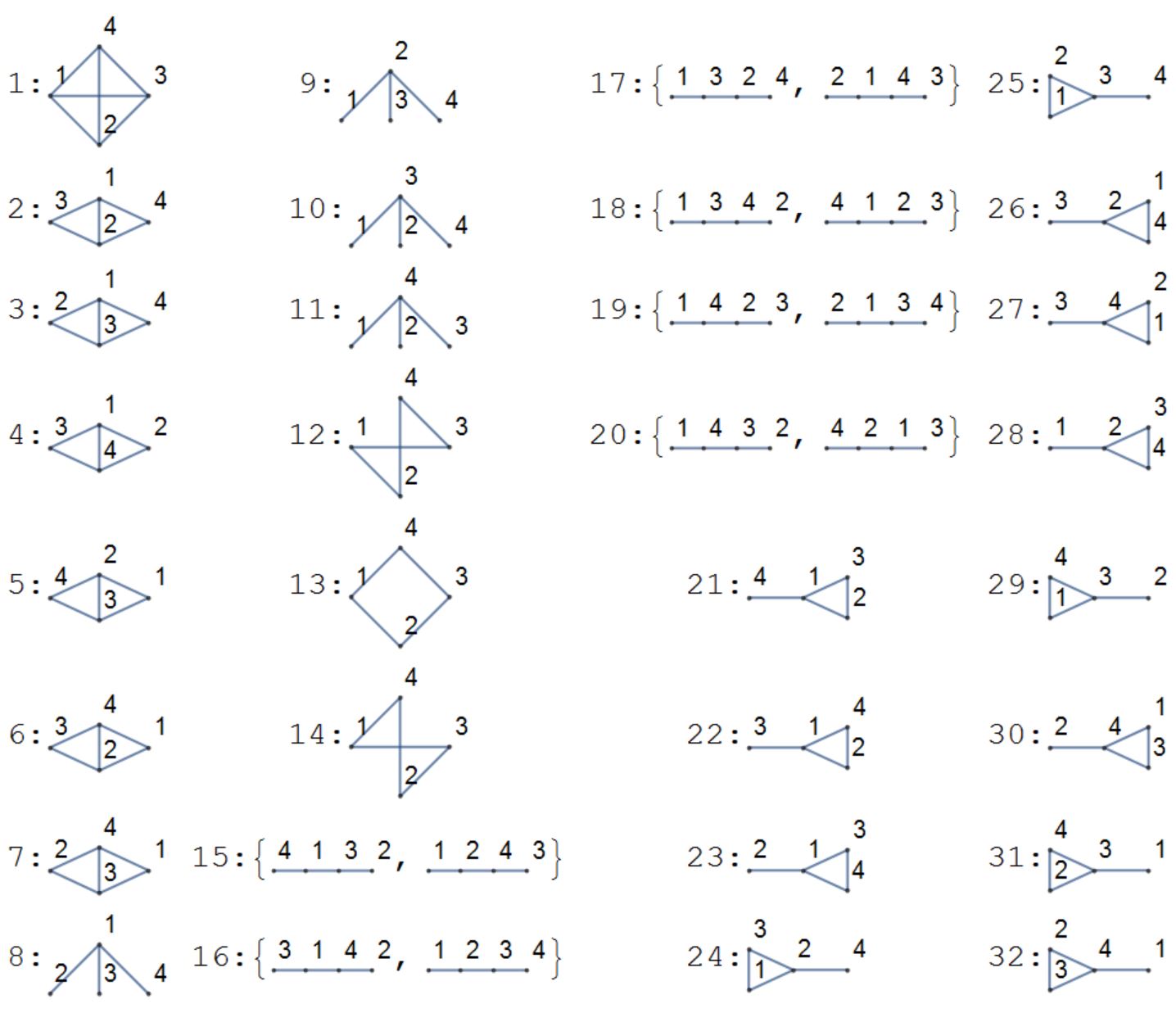}
\caption{There are 38 labeled simply connected graphs $G_i$ of order 4. Each $\mathrm{null}\left( W_{G_i,G_i}\right)$ compose 32 distinct subspaces with double degeneracy in the path graphs.}
\label{fig:g4iis}
\end{figure}

\begin{figure}[!htbp]
\centering
\includegraphics[scale=0.5]{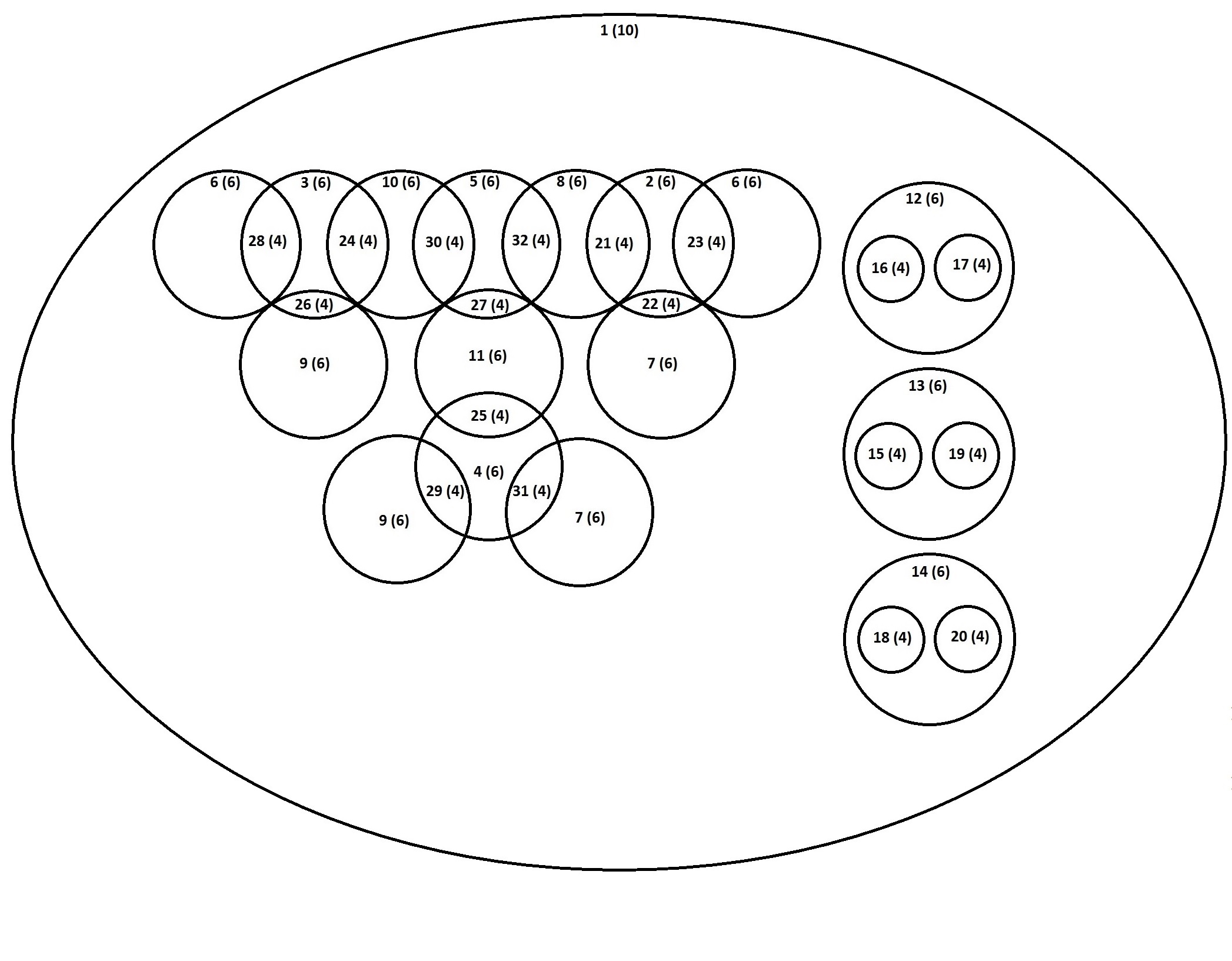}
\caption{Euler diagram for the 32 enumerated null spaces of $W_{G_i,G_i}$ in figure \ref{fig:g4iis}. The number of basis vectors spanning each null space is given in parentheses. (Only those intersections which are among the 32 distinct subspaces are shown.) Some labeled curves are not unique.} 
\label{fig:euler4ii}
\end{figure}

There are 38 different labeled simply connected graphs with order 4. When $i=j$, the null space of each $W_{G_i,G_i}$ compose 32 distinct subspaces with double degeneracy among the 12 labeled path graphs (see Fig. \ref{fig:g4iis}). An Euler diagram illustrating the interrelationships of the subspaces enumerated in Fig. \ref{fig:g4iis} is depicted in Fig. \ref{fig:euler4ii}.

There are an additional ${38 \choose 2} = 703$ possible $W_{G_i,G_j}$, where $i \neq j$, giving a total of 741 cases. All 741 possible $\mathrm{null}\left( W_{G_i,G_j}\right)$ comprise a total of 50 distinct null spaces. A summary of the cardinality and degeneracy for each null space is given in Table \ref{tab:W4ij}.

\section{Conclusion}

\begin{table}[ht]
\centering
\small

\begin{tabular}{|c|c|c|}
\hline
Zone & Degeneracy & Cardinality\\
\hline \hline
 1&1&10 \\
\hline \hline
 2&2&6 \\
\hline 
 3&2&6 \\
\hline 
 4&2&6 \\
\hline
 5&2&6 \\
\hline
 6&2&6 \\
\hline 
 7&2&6 \\
\hline 
 8&2&6 \\
\hline
 9&2&6 \\
\hline
 10&2&6 \\
\hline
 11&2&6 \\
\hline
 12&2&6 \\
\hline
 13&2&6 \\
\hline 
 14&2&6 \\
\hline \hline
 15&7&4 \\
\hline
 16&7&4 \\  
\hline
 17&7&4 \\  
\hline
 18&7&4 \\  
\hline
 19&7&4 \\  
\hline
 20&7&4 \\  
\hline \hline
 21&5&4 \\  
\hline 
 22&5&4 \\  
\hline
 23&5&4 \\  
\hline
 24&5&4 \\  
\hline
 25&5&4 \\  
\hline      
\end{tabular}
\quad
\begin{tabular}{|c|c|c|}
\hline
Zone & Degeneracy & Cardinality\\
\hline \hline
 26&5&4 \\
\hline 
 27&5&4 \\
\hline
 28&5&4 \\
\hline 
 29&5&4 \\
\hline
 30&5&4 \\
\hline
 31&5&4 \\
\hline 
 32&5&4 \\
\hline \hline
 33&3&4 \\
\hline
 34&3&4 \\
\hline
 35&3&4 \\
\hline
 36&3&4 \\
\hline \hline
 37&24&2 \\
\hline
 38&24&2 \\
\hline 
 39&24&2 \\
\hline \hline
 40&12&2 \\
\hline
 41&12&2 \\
\hline
 42&12&2 \\
\hline
 43&12&2 \\
\hline
 44&12&2 \\
\hline
 45&12&2 \\
\hline
 46&12&2 \\
\hline
 47&12&2 \\
\hline 
 48&12&2 \\
\hline
 49&12&2 \\
\hline \hline
 50&408&1 \\  
\hline    
\end{tabular}
\caption{The cardinality and degeneracy of the 50 distinct null spaces of $W_{G_i,G_j}$ for all 38 labeled graphs of order 4.}
\label{tab:W4ij}
\end{table}


\section*{References}
\bibliographystyle{unsrt}
\bibliography{Quantum_indistinguishability_of_graphs}

\end{document}